\documentclass[twocolumn,secnumarabic,amssymb, nobibnotes, aps, prl, superscriptaddress]{revtex4-2}

\usepackage{amsmath,mathrsfs,amssymb,amsfonts,amsthm,mathtools}
\usepackage{graphicx,color}
\usepackage{physics}
\usepackage[T1]{fontenc}
\usepackage{bm}
\usepackage{empheq}
\usepackage{hyperref}
\hypersetup{
    colorlinks=true,
    linkcolor=blue,
    citecolor=blue,
    urlcolor=blue
}

\newcommand{\nn}{\nonumber}
\newcommand*\widefbox[1]{\fbox{\hspace{2em}#1\hspace{2em}}}

\begin{document}
\title{Quantum Complementarity ad Infinitum: \\ Switching Higher-Order Coherence from Infinity to Zero}

\author{Arash Azizi}
\email{sazizi@tamu.edu}
\affiliation{The Institute for Quantum Science and Engineering, Texas A\&M University, College Station, TX 77843, U.S.A.}
\affiliation{Department of Physics and Astronomy, Texas A\&M University, College Station, TX 77843, U.S.A.}

\begin{abstract}
We report a profound manifestation of quantum complementarity in the higher-order photon statistics of the ``Janus state,'' a coherent superposition of two squeezed vacua. We find that the state acts as a perfect quantum switch for multi-photon correlations, toggled by the availability of which-path information. Erasing this information activates quantum interference that can be tuned to be maximally destructive. This reveals a remarkable hierarchy of suppression: while two-photon correlations remain finite, we prove analytically and demonstrate numerically that it is possible to drive all higher-order correlations ($g^{(k)}$ for $k \ge 3$) to zero. This transition from the extreme bunching of the constituent states ($g^{(k)} \to \infty$) to a state of profound quantum order is visualized by the emergence of negativity in the state's Wigner function, an unambiguous signature of non-classicality. This work provides a foundational demonstration of quantum complementarity in multi-photon statistics and introduces a new paradigm for engineering highly ordered, non-classical light from Gaussian resources.
\end{abstract}

\maketitle

\textit{Introduction.}---The principle of complementarity, first articulated by Bohr, establishes a fundamental trade-off between the wave-like and particle-like properties of a quantum system \cite{Bohr1928}. This concept, rooted in de Broglie's wave-particle duality \cite{debroglie1923}, has been formalized in modern quantum theory through inequalities that quantitatively link the visibility of interference fringes to the availability of ``which-path'' information \cite{Wootters1979, Scully1991, Englert1996, JAKOB2010Bergu}. While extensively demonstrated in single-particle interferometry, the manifestation of complementarity in the higher-order, multi-photon statistics of nonclassical light remains a rich and underexplored frontier.

The advent of quantum optics, guided by the coherence framework of Glauber and Sudarshan \cite{glauber1963coherent, sudarshan1963equivalence}, has enabled the generation of states like squeezed vacua, whose statistical properties defy classical description \cite{walls1983squeezed,slusher1985observatio,wu1986generation}. Recently, the ``Janus state,'' a coherent superposition of two such squeezed states, was shown to convert the intrinsic photon bunching of its components into strong two-photon antibunching ($g^{(2)}<1$) via quantum interference \cite{Azizi2025Janus}. This makes it an ideal platform to investigate how complementarity governs more complex, multi-photon correlations.

In this Letter, we extend this investigation to all higher-order coherences, $g^{(k)}$, and report a profound result: the Janus state acts as a perfect quantum switch for multi-photon statistics, toggled by the principle of complementarity. By deriving the exact $g^{(k)}$ function, we show that if which-path information is available—or if the superposition is configured for trivial interference—the system exhibits extreme photon bunching ($g^{(k)} \to \infty$), inheriting the divergent statistics of its constituent states. Conversely, if the path information is erased and the interference is non-trivial, a powerful mechanism to control the photon statistics emerges. We demonstrate that this interference can be tuned to be maximally destructive, creating the opportunity to completely suppress multi-photon events. Specifically, we show that while two-photon correlations converge to a finite value, it is possible to drive $g^{(k)} \to 0$ for all higher orders ($k \ge 3$).

We confirm this quantum switch with a set of rich analytical scaling laws that depend on the parity of $k$, and provide a direct visualization of the underlying nonclassicality via the negative regions of the state's Wigner function. This work provides a textbook demonstration of complementarity in a higher-order coherence context and establishes a new, experimentally feasible route to engineering quantum light with tailored multi-photon statistics for applications in quantum information and metrology \cite{weedbrook2012gaussian, asavanant2024multipartite}.
\begin{figure}[t]
    \centering
    \includegraphics[width=\columnwidth]{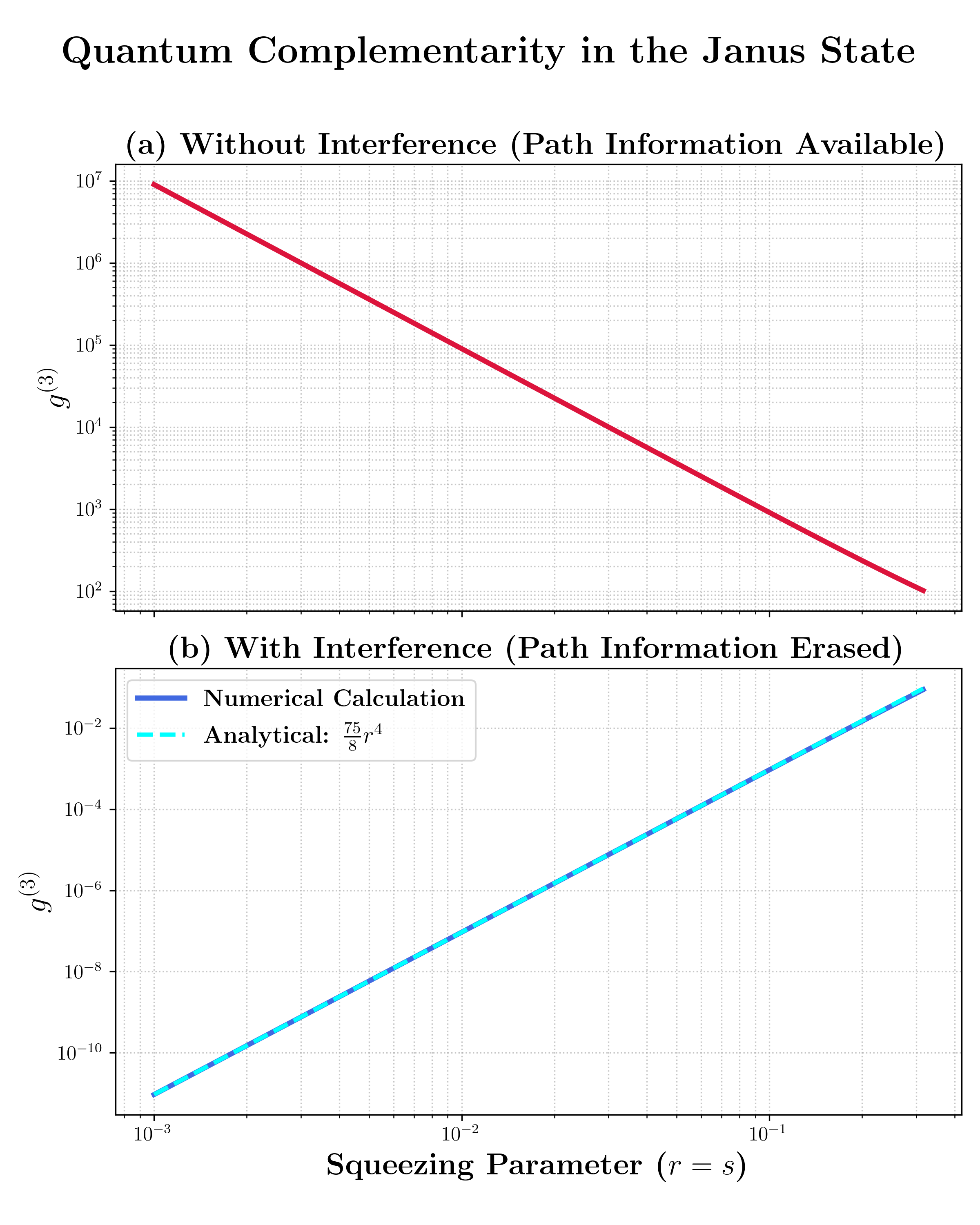}
    \caption{
        \textbf{The Complementarity Plot: Infinite vs. Zero Correlation.} (a) Without interference (path information available), the system behaves as a single squeezed state, whose $g^{(3)}$ diverges as the squeezing $r \to 0$. (b) With interference (path information erased), the Janus state exhibits profound three-photon suppression. This panel shows the special, optimal case where the interference phase is tuned to $\Delta=\pi$. In this configuration, a second-order destructive interference occurs, leading to a stronger suppression that scales as $g^{(3)} \propto r^4$, a special case of the general scaling laws. The numerical calculation (solid blue) perfectly matches this analytical prediction (dashed cyan), confirming that $g^{(3)}$ can be driven arbitrarily close to zero.
    }
    \label{fig:complementarity}
\end{figure}

\textit{Theory and Formalism.}---The Janus state is a normalized superposition of two squeezed vacua, $|\xi\rangle = |r e^{i\theta}\rangle$ and $|\zeta\rangle = |s e^{i\phi}\rangle$:
\begin{equation}
    |\psi\rangle = \chi |\xi\rangle + \eta |\zeta\rangle,
    \label{eq:superposition_main}
\end{equation}
where the complex amplitudes $\chi$ and $\eta$ are constrained by the normalization condition $\langle\psi|\psi\rangle=1$. Due to the non-orthogonality of the squeezed states, this implies:
\begin{equation}
   |\chi|^2 + |\eta|^2 + 2\mathscr{Re}\left[\chi\eta^* \frac{(1-x)^{1/4}(1-y)^{1/4}}{\sqrt{1-z}}\right] = 1.
   \label{eq:norm}
\end{equation}
This constraint, where $x = \tanh^2 r$, $y = \tanh^2 s$, and $z = \tanh r \tanh s e^{i\Delta}$, and $\Delta=\theta-\phi$, plays a central physical role in defining the allowed parameter space.

To analyze the multi-photon statistics, we study the higher-order coherence function:
\begin{equation}
    g^{(k)}(0) = \frac{\mathcal{N}_k}{\mathcal{N}_1^k}, \quad \text{where} \quad \mathcal{N}_k = \langle \psi| a^{\dagger k} a^k |\psi\rangle.
    \label{eq:gk_def}
\end{equation}
As detailed in the Supplemental Material (SM), we derive the moments $\mathcal{N}_k$ using a generating function formalism. This approach is built upon a family of what we term ``squeezing polynomials,'' $P_k(z)$, which obey a simple recurrence relation and fully determine the moments of the state. This method yields the exact analytical expression for the $k^\text{th}$ order moment:
\begin{align}
   \mathcal{N}_k =& |\chi|^2 \frac{P_k(x)}{(1-x)^k} + |\eta|^2 \frac{P_k(y)}{(1-y)^k}
   \nn\\
   &+ 2\mathscr{Re}\left[\chi \eta^* \frac{(1-x)^{1/4}(1-y)^{1/4} P_k(z)}{(1-z)^{k+1/2}}\right]. \label{eq:N_k_general}
\end{align}
The general analytical expression for the $k^\text{th}$-order coherence function is presented in Eq.~(\ref{g^k}) of SM. As a crucial validation of this general formalism, we consider the limit of a single squeezed state ($|\eta|=0, |\chi|=1$). Using Eq.~\eqref{eq:N_k_general} for $k=1$ and $k=3$, our expressions correctly yield the well-known result $g^{(3)} = 15 + 9/\sinh^2 r$, which diverges as $r \to 0$, establishing the extreme bunching shown in Fig.~\ref{fig:complementarity}(a).

\begin{figure}[t]
    \centering
    \includegraphics[width=\columnwidth]{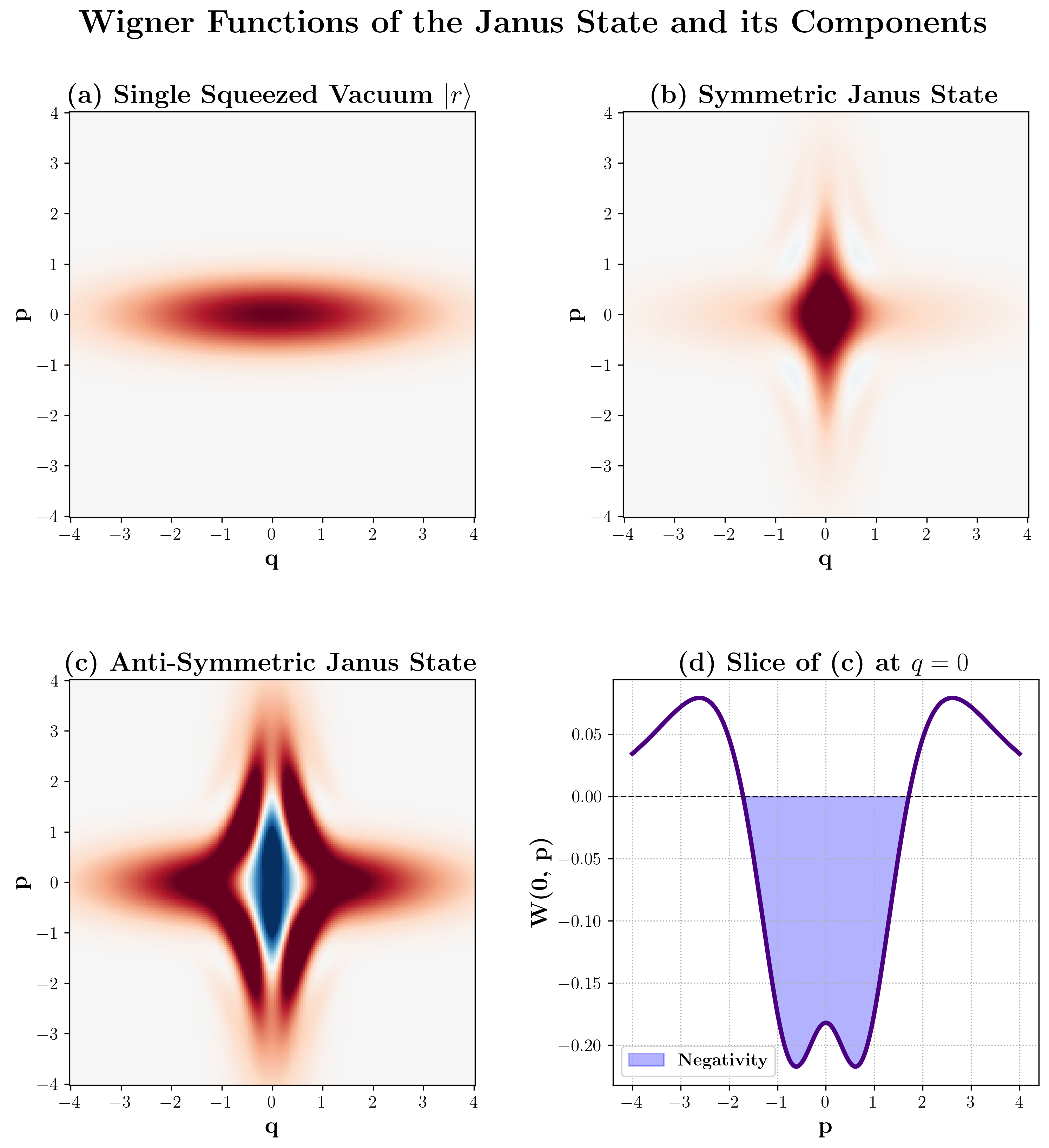}
    \caption{
        \textbf{Phase-Space Portrait of the Janus State.} (a) Wigner function of a single squeezed vacuum. (b) The symmetric Janus state shows constructive interference. (c) The anti-symmetric Janus state exhibits destructive interference, creating a region of negativity (blue) at its center---a hallmark of a non-classical state. (d) A slice through the center of (c) explicitly shows the negative values responsible for the suppression of multi-photon events.
    }
    \label{fig:wigner}
\end{figure}

\textit{Quantum Interference and Phase Space.}---The origin of the dramatic suppression of $g^{(3)}$ lies in the quantum interference between the two components of the Janus state. This can be visualized directly using the Wigner function, a quasi-probability distribution in phase space, shown in Fig.~\ref{fig:wigner}. A single squeezed state has a positive, elliptical Wigner function [Fig.~\ref{fig:wigner}(a)]. When two such states are superposed symmetrically, interference fringes appear, but the function remains positive [Fig.~\ref{fig:wigner}(b)], corresponding to a state with strong bunching.

However, for the anti-symmetric superposition that leads to suppression, the interference is destructive. This carves out a ``hole'' at the center of the phase space distribution where the Wigner function becomes negative [Fig.~\ref{fig:wigner}(c)]. The existence of this negativity is an unambiguous signature of a non-classical state and is the direct cause of the forbidden three-photon events. A slice through this region [Fig.~\ref{fig:wigner}(d)] explicitly shows the negative values and the oscillatory nature of the quantum interference.

\begin{figure}[t]
    \centering
    \includegraphics[width=\columnwidth]{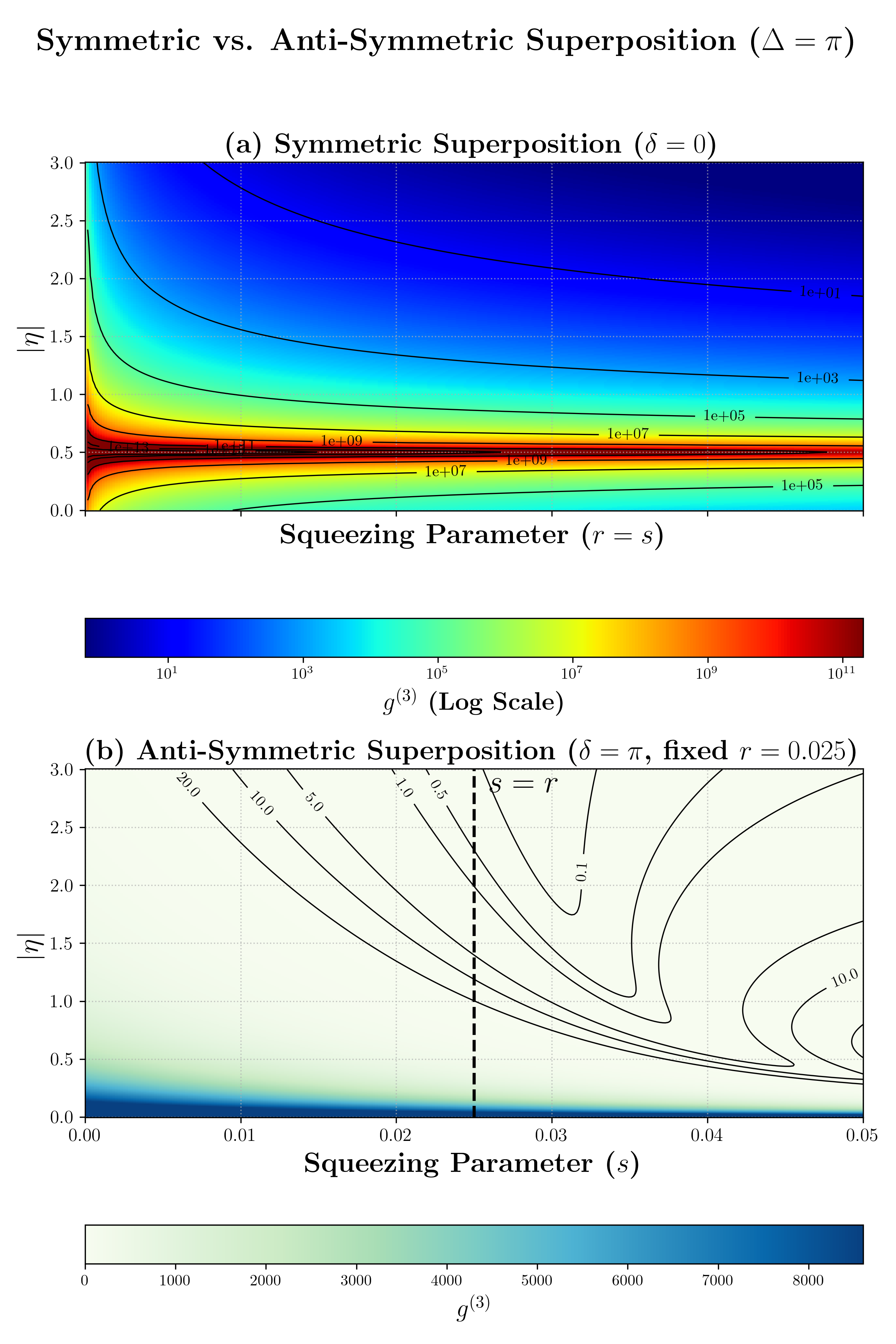}
    \caption{
        \textbf{Symmetric vs. Anti-Symmetric Superposition for $\Delta=\pi$.} (a) The symmetric case ($\delta=0$) shows strong bunching ($g^{(3)} \gg 1$). (b) The anti-symmetric case ($\delta=\pi$) reveals a rich landscape where $g^{(3)}$ can be suppressed below unity.
    }
    \label{fig:sym_vs_asym}
\end{figure}

\textit{Tunability and Control.}---The switch between bunching and antibunching is controlled by the relative phases of the superposition. Figure \ref{fig:sym_vs_asym} compares the two extreme cases for the interference phase $\Delta=\pi$. The symmetric superposition ($\delta=0$) exhibits strong bunching for all parameters [Fig.~\ref{fig:sym_vs_asym}(a)]. In contrast, the anti-symmetric superposition ($\delta=\pi$) shows a rich structure where $g^{(3)}$ can be suppressed [Fig.~\ref{fig:sym_vs_asym}(b)].

\begin{figure}[t]
    \centering
    \includegraphics[width=\columnwidth]{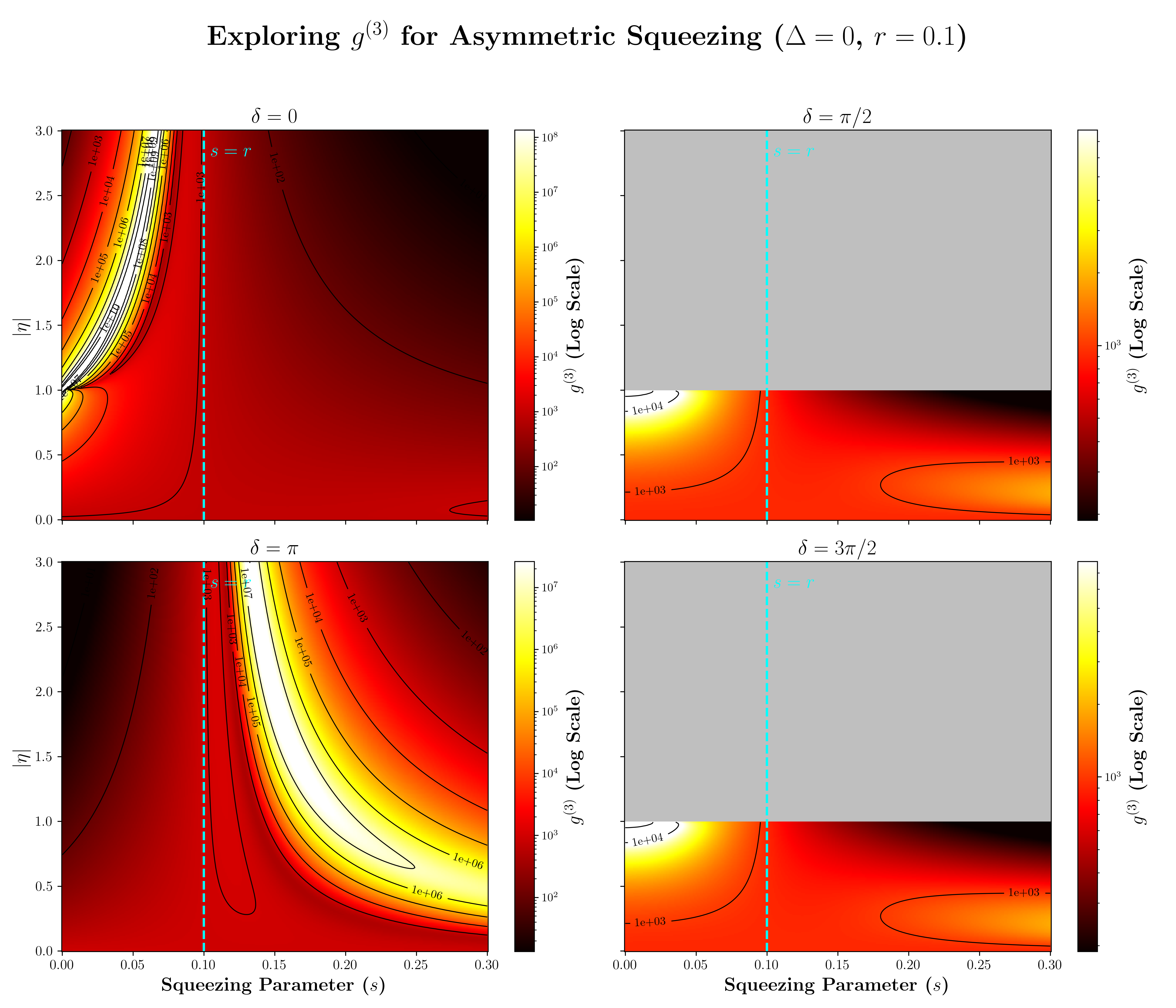}
    \caption{
        \textbf{Absence of Suppression for $\Delta=0$.} This figure explores the four branches of the superposition phase $\delta$ for the trivial interference case, $\Delta=0$. In all panels, the system exhibits strong photon bunching ($g^{(3)} \gg 1$), confirming that non-trivial interference is necessary for suppression.
    }
    \label{fig:delta_zero}
\end{figure}

However, this control is only possible when the interference is non-trivial. Figure \ref{fig:delta_zero} explores the case where the interference phase is trivial, $\Delta=0$. As shown, for all choices of the superposition phase $\delta$, the system remains locked in a regime of extreme photon bunching. A detailed analysis in the SM proves that the normalization constraint forbids access to the large-amplitude limit required for suppression when $\Delta=0$. This provides definitive proof that non-trivial interference ($\Delta \neq 0$) is an essential ingredient for observing the suppression of higher-order coherence.

\textit{Generalization to Higher-Order Coherence.}---The profound suppression of three-photon events is not an isolated phenomenon but a general feature of the Janus state's higher-order coherence. The formalism developed in the SM allows us to analyze the asymptotic behavior of the $k^{\text{th}}$ order coherence function, $g^{(k)}$, in the limit of small squeezing ($r \to 0$) with critically tuned amplitude ($|\eta| \approx 1/r$).

The scaling of $g^{(k)}$ critically depends on the interference phase $\Delta$ and whether $k$ is odd or even, yielding four distinct scenarios. In the generic case ($1 - \cos(l_k \Delta) \neq 0$, with $l_k = \lceil k/2 \rceil$):
\begin{equation}
    g^{(k)}(\psi) \propto 
    \begin{cases}
        r^{k-1} & \text{for odd } k, \\
        r^{k-2} & \text{for even } k.
    \end{cases}
    \label{eq:scaling_law_general}
\end{equation}
In the special case ($1 - \cos(l_k \Delta) = 0$, i.e., $l_k \Delta = 2n\pi$), higher-order terms lead to stronger suppression:
\begin{equation}
    g^{(k)}(\psi) \propto 
    \begin{cases}
        r^{k+1} & \text{for odd } k, \\
        r^{k} & \text{for even } k.
    \end{cases}
\end{equation}

\begin{figure}[t]
    \centering
    \includegraphics[width=\columnwidth]{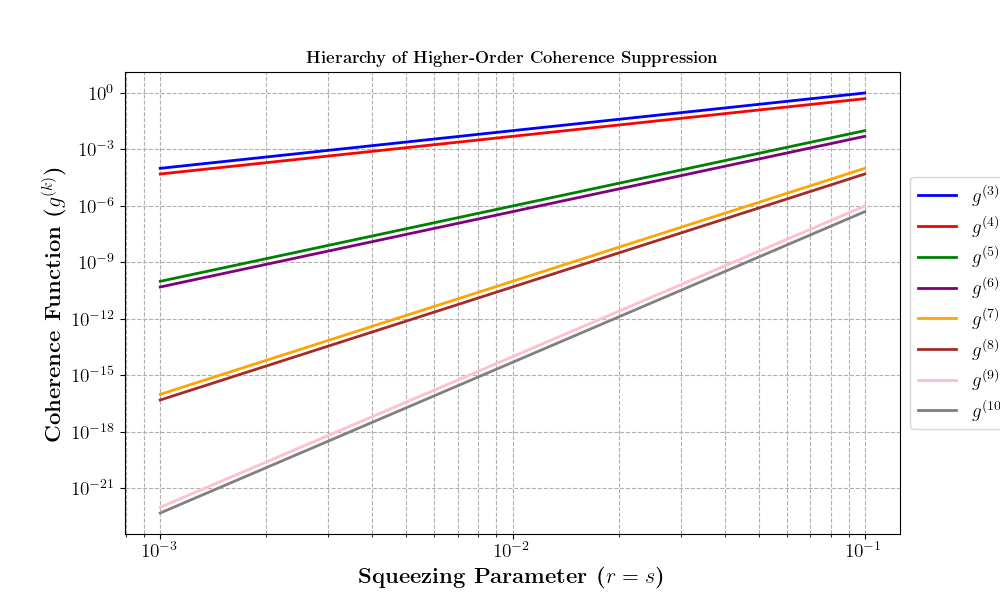}
    \caption{
        \textbf{Hierarchy of Higher-Order Coherence Suppression.} Log-log plot of $g^{(k)}$ vs. squeezing parameter $r$. While $g^{(2)}$ (not shown) flattens to a constant, all higher orders ($k \ge 3$) are suppressed to zero, with stronger suppression for higher $k$ due to increasing exponents (paired for odd-even $k$ in the generic case). In special phase cases, the decay is even faster (e.g., $\propto r^{k+1}$ for odd $k$, $\propto r^{k}$ for even $k$; not shown).
    }
    \label{fig:hierarchy}
\end{figure}

This hierarchy in the photon statistics is illustrated in Fig.~\ref{fig:hierarchy}. For second-order coherence ($k=2$), the generic scaling is $r^{0}$, converging to a finite non-classical value ($\approx 0.567$ \cite{Azizi2025Janus}). For $k \ge 3$, the exponents are positive, implying $g^{(k)} \to 0$, with even stronger decay in the special phase configuration for specific $k$.

This generalization strengthens the complementarity argument: quantum interference in the Janus state eliminates multi-photon bunching beyond pairwise correlations, toggling from divergent higher-order correlations in the single squeezed state limit to a highly ordered non-classical photon stream.

\textit{Experimental Feasibility.}---The generation and characterization of the Janus state are well within the capabilities of modern quantum optics laboratories. The constituent single-mode squeezed vacuum states are routinely generated using optical parametric oscillators (OPOs) operating below threshold, with experimentally demonstrated squeezing levels far exceeding what is required for our proposed effects \cite{vahlbruch2016detection}. The coherent superposition of these two squeezed modes can be achieved deterministically using a phase-stable interferometer, a technique that has recently been perfected for creating complex non-Gaussian states from squeezed-state inputs. The precise control over the relative phases ($\Delta, \delta$) and amplitudes ($\chi, \eta$) required to navigate the phase landscape can be implemented with standard, high-speed electro-optic modulators and piezo-controlled mirrors. Finally, the resulting higher-order photon statistics can be directly measured using Hanbury Brown--Twiss-style setups equipped with state-of-the-art photon-number-resolving detectors, such as superconducting nanowire single-photon detectors (SNSPDs), which now offer near-unity quantum efficiency \cite{Zadeh2021}. While experimental imperfections such as photon loss will affect the purity of the state, our analysis shows that the core phenomena of tunable bunching and antibunching are robust and observable with moderate, readily achievable squeezing parameters.
 
\textit{Conclusion.}---In this Letter, we have investigated the higher-order quantum statistics of the Janus state, a superposition of two squeezed vacua. By deriving the exact coherence function for arbitrary order $k$, we have uncovered a profound manifestation of quantum complementarity. We have shown that the Janus state acts as a perfect quantum switch for multi-photon statistics: when which-path information is available, the state exhibits extreme photon bunching ($g^{(k)} \to \infty$), while erasing this information enables quantum interference that can be tuned to be maximally destructive.

This interference leads to our central result: the complete suppression of all higher-order correlations. We have proven analytically and demonstrated numerically that while two-photon correlations converge to a finite, non-classical value, it is possible to drive $g^{(k)} \to 0$ for all orders $k \ge 3$. This remarkable hierarchy of suppression, governed by elegant scaling laws, is a direct consequence of the non-classical nature of the state, which we have visualized through the negative regions of its Wigner function. This work establishes the Janus state as a powerful and experimentally accessible platform for exploring fundamental quantum principles and for engineering highly ordered, non-classical light with tailored multi-photon statistics for advanced quantum technologies.

\textit{Acknowledgments.}---I am grateful to Girish Agarwal, Marlan Scully, Bill Unruh, and Suhail Zubairy for discussions. This work was supported by the Robert A. Welch Foundation (Grant No. A-1261) and the National Science Foundation (Grant No. PHY-2013771).



\bibliographystyle{apsrev4-2}
\bibliography{SqueezingRef}

\begin{thebibliography}{16}%
\makeatletter
\providecommand \@ifxundefined [1]{%
 \@ifx{#1\undefined}
}%
\providecommand \@ifnum [1]{%
 \ifnum #1\expandafter \@firstoftwo
 \else \expandafter \@secondoftwo
 \fi
}%
\providecommand \@ifx [1]{%
 \ifx #1\expandafter \@firstoftwo
 \else \expandafter \@secondoftwo
 \fi
}%
\providecommand \natexlab [1]{#1}%
\providecommand \enquote  [1]{``#1''}%
\providecommand \bibnamefont  [1]{#1}%
\providecommand \bibfnamefont [1]{#1}%
\providecommand \citenamefont [1]{#1}%
\providecommand \href@noop [0]{\@secondoftwo}%
\providecommand \href [0]{\begingroup \@sanitize@url \@href}%
\providecommand \@href[1]{\@@startlink{#1}\@@href}%
\providecommand \@@href[1]{\endgroup#1\@@endlink}%
\providecommand \@sanitize@url [0]{\catcode `\\12\catcode `\$12\catcode `\&12\catcode `\#12\catcode `\^12\catcode `\_12\catcode `\%12\relax}%
\providecommand \@@startlink[1]{}%
\providecommand \@@endlink[0]{}%
\providecommand \url  [0]{\begingroup\@sanitize@url \@url }%
\providecommand \@url [1]{\endgroup\@href {#1}{\urlprefix }}%
\providecommand \urlprefix  [0]{URL }%
\providecommand \Eprint [0]{\href }%
\providecommand \doibase [0]{https://doi.org/}%
\providecommand \selectlanguage [0]{\@gobble}%
\providecommand \bibinfo  [0]{\@secondoftwo}%
\providecommand \bibfield  [0]{\@secondoftwo}%
\providecommand \translation [1]{[#1]}%
\providecommand \BibitemOpen [0]{}%
\providecommand \bibitemStop [0]{}%
\providecommand \bibitemNoStop [0]{.\EOS\space}%
\providecommand \EOS [0]{\spacefactor3000\relax}%
\providecommand \BibitemShut  [1]{\csname bibitem#1\endcsname}%
\let\auto@bib@innerbib\@empty
\bibitem [{\citenamefont {Bohr}(1928)}]{Bohr1928}%
  \BibitemOpen
  \bibfield  {author} {\bibinfo {author} {\bibfnamefont {N.}~\bibnamefont {Bohr}},\ }\href {https://doi.org/10.1038/121580a0} {\bibfield  {journal} {\bibinfo  {journal} {Nature}\ }\textbf {\bibinfo {volume} {121}},\ \bibinfo {pages} {580} (\bibinfo {year} {1928})}\BibitemShut {NoStop}%
\bibitem [{\citenamefont {de~Broglie}(1923)}]{debroglie1923}%
  \BibitemOpen
  \bibfield  {author} {\bibinfo {author} {\bibfnamefont {L.}~\bibnamefont {de~Broglie}},\ }\href {https://doi.org/10.1038/112540a0} {\bibfield  {journal} {\bibinfo  {journal} {Nature}\ }\textbf {\bibinfo {volume} {112}},\ \bibinfo {pages} {540} (\bibinfo {year} {1923})}\BibitemShut {NoStop}%
\bibitem [{\citenamefont {Wootters}\ and\ \citenamefont {Zurek}(1979)}]{Wootters1979}%
  \BibitemOpen
  \bibfield  {author} {\bibinfo {author} {\bibfnamefont {W.~K.}\ \bibnamefont {Wootters}}\ and\ \bibinfo {author} {\bibfnamefont {W.~H.}\ \bibnamefont {Zurek}},\ }\href {https://doi.org/10.1103/PhysRevD.19.473} {\bibfield  {journal} {\bibinfo  {journal} {Phys. Rev. D}\ }\textbf {\bibinfo {volume} {19}},\ \bibinfo {pages} {473} (\bibinfo {year} {1979})}\BibitemShut {NoStop}%
\bibitem [{\citenamefont {Scully}\ \emph {et~al.}(1991)\citenamefont {Scully}, \citenamefont {Englert},\ and\ \citenamefont {Walther}}]{Scully1991}%
  \BibitemOpen
  \bibfield  {author} {\bibinfo {author} {\bibfnamefont {M.~O.}\ \bibnamefont {Scully}}, \bibinfo {author} {\bibfnamefont {B.-G.}\ \bibnamefont {Englert}},\ and\ \bibinfo {author} {\bibfnamefont {H.}~\bibnamefont {Walther}},\ }\href {https://doi.org/10.1038/351111a0} {\bibfield  {journal} {\bibinfo  {journal} {Nature}\ }\textbf {\bibinfo {volume} {351}},\ \bibinfo {pages} {111} (\bibinfo {year} {1991})}\BibitemShut {NoStop}%
\bibitem [{\citenamefont {Englert}(1996)}]{Englert1996}%
  \BibitemOpen
  \bibfield  {author} {\bibinfo {author} {\bibfnamefont {B.-G.}\ \bibnamefont {Englert}},\ }\href {https://doi.org/10.1103/PhysRevLett.77.2154} {\bibfield  {journal} {\bibinfo  {journal} {Phys. Rev. Lett.}\ }\textbf {\bibinfo {volume} {77}},\ \bibinfo {pages} {2154} (\bibinfo {year} {1996})}\BibitemShut {NoStop}%
\bibitem [{\citenamefont {Jakob}\ and\ \citenamefont {Bergou}(2010)}]{JAKOB2010Bergu}%
  \BibitemOpen
  \bibfield  {author} {\bibinfo {author} {\bibfnamefont {M.}~\bibnamefont {Jakob}}\ and\ \bibinfo {author} {\bibfnamefont {J.~A.}\ \bibnamefont {Bergou}},\ }\href {https://doi.org/https://doi.org/10.1016/j.optcom.2009.10.044} {\bibfield  {journal} {\bibinfo  {journal} {Optics Communications}\ }\textbf {\bibinfo {volume} {283}},\ \bibinfo {pages} {827} (\bibinfo {year} {2010})}\BibitemShut {NoStop}%
\bibitem [{\citenamefont {Glauber}(1963)}]{glauber1963coherent}%
  \BibitemOpen
  \bibfield  {author} {\bibinfo {author} {\bibfnamefont {R.~J.}\ \bibnamefont {Glauber}},\ }\href {https://doi.org/10.1103/PhysRev.131.2766} {\bibfield  {journal} {\bibinfo  {journal} {Phys. Rev.}\ }\textbf {\bibinfo {volume} {131}},\ \bibinfo {pages} {2766} (\bibinfo {year} {1963})}\BibitemShut {NoStop}%
\bibitem [{\citenamefont {Sudarshan}(1963)}]{sudarshan1963equivalence}%
  \BibitemOpen
  \bibfield  {author} {\bibinfo {author} {\bibfnamefont {E.~C.~G.}\ \bibnamefont {Sudarshan}},\ }\href {https://doi.org/10.1103/PhysRevLett.10.277} {\bibfield  {journal} {\bibinfo  {journal} {Phys. Rev. Lett.}\ }\textbf {\bibinfo {volume} {10}},\ \bibinfo {pages} {277} (\bibinfo {year} {1963})}\BibitemShut {NoStop}%
\bibitem [{\citenamefont {Walls}(1983)}]{walls1983squeezed}%
  \BibitemOpen
  \bibfield  {author} {\bibinfo {author} {\bibfnamefont {D.~F.}\ \bibnamefont {Walls}},\ }\href@noop {} {\bibfield  {journal} {\bibinfo  {journal} {nature}\ }\textbf {\bibinfo {volume} {306}},\ \bibinfo {pages} {141} (\bibinfo {year} {1983})}\BibitemShut {NoStop}%
\bibitem [{\citenamefont {Slusher}\ \emph {et~al.}(1985)\citenamefont {Slusher}, \citenamefont {Hollberg}, \citenamefont {Yurke}, \citenamefont {Mertz},\ and\ \citenamefont {Valley}}]{slusher1985observatio}%
  \BibitemOpen
  \bibfield  {author} {\bibinfo {author} {\bibfnamefont {R.~E.}\ \bibnamefont {Slusher}}, \bibinfo {author} {\bibfnamefont {L.~W.}\ \bibnamefont {Hollberg}}, \bibinfo {author} {\bibfnamefont {B.}~\bibnamefont {Yurke}}, \bibinfo {author} {\bibfnamefont {J.~C.}\ \bibnamefont {Mertz}},\ and\ \bibinfo {author} {\bibfnamefont {J.~F.}\ \bibnamefont {Valley}},\ }\href {https://doi.org/10.1103/PhysRevLett.55.2409} {\bibfield  {journal} {\bibinfo  {journal} {Phys. Rev. Lett.}\ }\textbf {\bibinfo {volume} {55}},\ \bibinfo {pages} {2409} (\bibinfo {year} {1985})}\BibitemShut {NoStop}%
\bibitem [{\citenamefont {Wu}\ \emph {et~al.}(1986)\citenamefont {Wu}, \citenamefont {Kimble}, \citenamefont {Hall},\ and\ \citenamefont {Wu}}]{wu1986generation}%
  \BibitemOpen
  \bibfield  {author} {\bibinfo {author} {\bibfnamefont {L.-A.}\ \bibnamefont {Wu}}, \bibinfo {author} {\bibfnamefont {H.~J.}\ \bibnamefont {Kimble}}, \bibinfo {author} {\bibfnamefont {J.~L.}\ \bibnamefont {Hall}},\ and\ \bibinfo {author} {\bibfnamefont {H.}~\bibnamefont {Wu}},\ }\href {https://doi.org/10.1103/PhysRevLett.57.2520} {\bibfield  {journal} {\bibinfo  {journal} {Phys. Rev. Lett.}\ }\textbf {\bibinfo {volume} {57}},\ \bibinfo {pages} {2520} (\bibinfo {year} {1986})}\BibitemShut {NoStop}%
\bibitem [{\citenamefont {Azizi}(2025)}]{Azizi2025Janus}%
  \BibitemOpen
  \bibfield  {author} {\bibinfo {author} {\bibfnamefont {A.}~\bibnamefont {Azizi}},\ }\href {https://arxiv.org/abs/2506.06397} {\bibinfo {title} {{The Janus State: Strong Photon Antibunching from a Superposition of Squeezed Vacua}}} (\bibinfo {year} {2025}),\ \Eprint {https://arxiv.org/abs/2506.06397} {arXiv:2506.06397 [quant-ph]} \BibitemShut {NoStop}%
\bibitem [{\citenamefont {Weedbrook}\ \emph {et~al.}(2012)\citenamefont {Weedbrook}, \citenamefont {Pirandola}, \citenamefont {Garc\'{\i}a-Patr\'on}, \citenamefont {Cerf}, \citenamefont {Ralph}, \citenamefont {Shapiro},\ and\ \citenamefont {Lloyd}}]{weedbrook2012gaussian}%
  \BibitemOpen
  \bibfield  {author} {\bibinfo {author} {\bibfnamefont {C.}~\bibnamefont {Weedbrook}}, \bibinfo {author} {\bibfnamefont {S.}~\bibnamefont {Pirandola}}, \bibinfo {author} {\bibfnamefont {R.}~\bibnamefont {Garc\'{\i}a-Patr\'on}}, \bibinfo {author} {\bibfnamefont {N.~J.}\ \bibnamefont {Cerf}}, \bibinfo {author} {\bibfnamefont {T.~C.}\ \bibnamefont {Ralph}}, \bibinfo {author} {\bibfnamefont {J.~H.}\ \bibnamefont {Shapiro}},\ and\ \bibinfo {author} {\bibfnamefont {S.}~\bibnamefont {Lloyd}},\ }\href {https://doi.org/10.1103/RevModPhys.84.621} {\bibfield  {journal} {\bibinfo  {journal} {Rev. Mod. Phys.}\ }\textbf {\bibinfo {volume} {84}},\ \bibinfo {pages} {621} (\bibinfo {year} {2012})}\BibitemShut {NoStop}%
\bibitem [{\citenamefont {Asavanant}\ and\ \citenamefont {Furusawa}(2024)}]{asavanant2024multipartite}%
  \BibitemOpen
  \bibfield  {author} {\bibinfo {author} {\bibfnamefont {W.}~\bibnamefont {Asavanant}}\ and\ \bibinfo {author} {\bibfnamefont {A.}~\bibnamefont {Furusawa}},\ }\href {https://doi.org/10.1103/PhysRevA.109.040101} {\bibfield  {journal} {\bibinfo  {journal} {Phys. Rev. A}\ }\textbf {\bibinfo {volume} {109}},\ \bibinfo {pages} {040101} (\bibinfo {year} {2024})}\BibitemShut {NoStop}%
\bibitem [{\citenamefont {Vahlbruch}\ \emph {et~al.}(2016)\citenamefont {Vahlbruch}, \citenamefont {Mehmet}, \citenamefont {Danzmann},\ and\ \citenamefont {Schnabel}}]{vahlbruch2016detection}%
  \BibitemOpen
  \bibfield  {author} {\bibinfo {author} {\bibfnamefont {H.}~\bibnamefont {Vahlbruch}}, \bibinfo {author} {\bibfnamefont {M.}~\bibnamefont {Mehmet}}, \bibinfo {author} {\bibfnamefont {K.}~\bibnamefont {Danzmann}},\ and\ \bibinfo {author} {\bibfnamefont {R.}~\bibnamefont {Schnabel}},\ }\href {https://doi.org/10.1103/PhysRevLett.117.110801} {\bibfield  {journal} {\bibinfo  {journal} {Phys. Rev. Lett.}\ }\textbf {\bibinfo {volume} {117}},\ \bibinfo {pages} {110801} (\bibinfo {year} {2016})}\BibitemShut {NoStop}%
\bibitem [{\citenamefont {Esmaeil~Zadeh}\ \emph {et~al.}(2021)\citenamefont {Esmaeil~Zadeh}, \citenamefont {Chang}, \citenamefont {Los}, \citenamefont {Gyger}, \citenamefont {Elshaari}, \citenamefont {Steinhauer}, \citenamefont {Dorenbos},\ and\ \citenamefont {Zwiller}}]{Zadeh2021}%
  \BibitemOpen
  \bibfield  {author} {\bibinfo {author} {\bibfnamefont {I.}~\bibnamefont {Esmaeil~Zadeh}}, \bibinfo {author} {\bibfnamefont {J.}~\bibnamefont {Chang}}, \bibinfo {author} {\bibfnamefont {J.~W.~N.}\ \bibnamefont {Los}}, \bibinfo {author} {\bibfnamefont {S.}~\bibnamefont {Gyger}}, \bibinfo {author} {\bibfnamefont {A.~W.}\ \bibnamefont {Elshaari}}, \bibinfo {author} {\bibfnamefont {S.}~\bibnamefont {Steinhauer}}, \bibinfo {author} {\bibfnamefont {S.~N.}\ \bibnamefont {Dorenbos}},\ and\ \bibinfo {author} {\bibfnamefont {V.}~\bibnamefont {Zwiller}},\ }\href {https://doi.org/10.1063/5.0045990} {\bibfield  {journal} {\bibinfo  {journal} {Applied Physics Letters}\ }\textbf {\bibinfo {volume} {118}},\ \bibinfo {pages} {190502} (\bibinfo {year} {2021})},\ \Eprint {https://arxiv.org/abs/https://pubs.aip.org/aip/apl/article-pdf/doi/10.1063/5.0045990/20021815/190502\_1\_5.0045990.pdf} {https://pubs.aip.org/aip/apl/article-pdf/doi/10.1063/5.0045990/20021815/190502\_1\_5.0045990.pdf} \BibitemShut {NoStop}%
\end{thebibliography}%

\newpage
\clearpage
\onecolumngrid
\appendix
\section*{Supplemental Material for: ``Quantum Complementarity ad Infinitum: Switching Higher-Order Coherence from Infinity to Zero''}
\author{Arash Azizi}
\affiliation{The Institute for Quantum Science and Engineering, Texas A\&M University, College Station, TX 77843, U.S.A.}
\affiliation{Department of Physics and Astronomy, Texas A\&M University, College Station, TX 77843, U.S.A.}
\date{\today}
\maketitle
\setcounter{equation}{0}
\renewcommand{\theequation}{S\arabic{equation}}

In this Supplemental Material, we provide a detailed, pedagogical derivation of the higher-order coherence properties of the Janus state. We begin by establishing a general formalism for the $k^{\text{th}}$ order moments of a superposition of two squeezed states, introducing the ``squeezing polynomials'' that govern their behavior. We then apply this formalism to derive the explicit expressions for $g^{(k)}$ and analyze the asymptotic limits that reveal the extreme tunability of the system's photon statistics.

\section*{Calculating \texorpdfstring{$\langle \psi | a^{\dagger k} a^k | \psi \rangle$}{}}
We begin by defining the unnormalized Janus state as a superposition of two single-mode squeezed vacua, $|\xi\rangle = |re^{i\theta}\rangle$ and $|\zeta\rangle = |se^{i\phi}\rangle$:
\begin{align}
    |\psi\rangle &= \chi|\xi\rangle + \eta|\zeta\rangle,
\end{align}
where $\chi$ and $\eta$ are complex amplitudes. The squeezed vacuum states are expressed in the Fock basis as:
\begin{align}
    |\xi\rangle &= (1-|\alpha|^2)^{1/4} \sum_{n=0}^{\infty} \frac{\sqrt{(2n)!}}{n!} \left(-\frac{\alpha}{2}\right)^n |2n\rangle, \quad \text{where } \alpha = -\tanh r \, e^{i\theta}, \nn\\
    |\zeta\rangle &= (1-|\beta|^2)^{1/4} \sum_{m=0}^{\infty} \frac{\sqrt{(2m)!}}{m!} \left(-\frac{\beta}{2}\right)^m |2m\rangle, \quad \text{where } \beta = -\tanh s \, e^{i\phi}.
\end{align}
The $k^{\text{th}}$ order correlation function is defined as:
\begin{align}
    g^{(k)}(0) &= \frac{\langle a^{\dagger k} a^k \rangle}{\langle a^\dagger a \rangle^k} = \frac{\langle \psi | a^{\dagger k} a^k | \psi \rangle}{\langle \psi | a^\dagger a | \psi \rangle^k} \equiv \frac{\mathcal{N}_k}{\mathcal{N}_1^k}
\end{align}
The numerator $\mathcal{N}_k$ expands to:
\begin{align}
    \mathcal{N}_k = |\chi|^2 \langle \xi | a^{\dagger k} a^k | \xi \rangle + |\eta|^2 \langle \zeta | a^{\dagger k} a^k | \zeta \rangle + 2\mathscr{Re} \left[ \chi\eta^* \langle \zeta | a^{\dagger k} a^k | \xi \rangle \right]
\end{align}
Our task is to find a general expression for the matrix element $\langle \zeta | a^{\dagger k} a^k | \xi \rangle$.

\subsection{A Generating Function for Matrix Elements}
We first calculate the inner product of the states produced by applying $a^k$ to each squeezed vacuum. The action of $a^k$ is:
\begin{align}
    a^k|\xi\rangle =& (1-|\alpha|^2)^{1/4} \sum_{n=0}^{\infty} 
     \left(-\frac{\alpha}{2}\right)^n \frac{1}{n!}
    \sqrt{(2n)!} \, \sqrt{(2n)(2n-1)...(2n-k+1)} \, |2n-k\rangle \nn\\
     =& (1-|\alpha|^2)^{1/4} \sum_{n=0}^{\infty} 
     \left(-\alpha\right)^n \frac{1}{2^n\,n!}
    \sqrt{(2n)!} \, \sqrt{\frac{(2n)!}{(2n-k)!}}\, |2n-k\rangle \nn\\
   =& (1-|\alpha|^2)^{1/4} \sum_{n=0}^{\infty} 
     \left(-\alpha\right)^n \frac{(2n)!!(2n-1)!!}{2^n\,n!\,\sqrt{(2n-k)!}}
    \, |2n-k\rangle \nn\\
   =& (1-|\alpha|^2)^{1/4} \sum_{n=0}^{\infty} 
     \left(-\alpha\right)^n \frac{(2n-1)!!}{\sqrt{(2n-k)!}}
    \, |2n-k\rangle,
\end{align}
where we have used $(2n)!=(2n)!!(2n-1)!!$, where $(2n)!!=2n(2n-2)\cdots 2$. 
Therefore we have
\begin{align}
 \boxed{\quad   a^k|\xi\rangle
 =(1-|\alpha|^2)^{1/4} \sum_{n=0}^{\infty} 
     \left(-\alpha\right)^n \frac{(2n-1)!!}{\sqrt{(2n-k)!}}
    \, |2n-k\rangle. \quad}
\end{align}
The inner product $\langle a^k \zeta | a^k \xi \rangle = \langle \zeta | a^{\dagger k} a^k | \xi \rangle$ becomes:
\begin{align}
    \langle \zeta& | (a^\dagger)^k a^k | \xi \rangle \nn\\
    &= (1-|\alpha|^2)^{1/4} (1-|\beta|^2)^{1/4} \sum_{m,n=0}^{\infty} \frac{(2m-1)!!}{\sqrt{(2m-k)!}} \frac{(2n-1)!!}{\sqrt{(2n-k)!}}  
    (-\beta^*)^m (-\alpha)^n \langle 2m-k | 2n-k \rangle \nn\\
    &= (1-|\alpha|^2)^{1/4} (1-|\beta|^2)^{1/4} \sum_{n=0}^{\infty} \frac{\left((2n-1)!!\right)^2 }{(2n-k)!}(\alpha \beta^*)^n \nn\\
    &= (1-|\alpha|^2)^{1/4} (1-|\beta|^2)^{1/4} \sum_{n=0}^{\infty} 
    \frac{(2n-1)!! }{(2n)!!} \frac{(2n)! }{(2n-k)!}(\alpha \beta^*)^n.
\end{align}
Let $z = \alpha \beta^*$. We can define a generating function $F_k(z)$ for the series:
\begin{align}
    F_k(z) &\equiv \sum_{n=\lceil k/2 \rceil}^{\infty} \frac{(2n)!}{(2n-k)!} \frac{(2n-1)!!}{(2n)!!} z^n
\end{align}
For $k=0$, we have $F_0(z) = \sum_{n=0}^{\infty} \frac{(2n-1)!!}{(2n)!!} z^n = (1-z)^{-1/2}$. For higher $k$, the function can be generated by applying a differential operator. Noting that the term $\frac{(2n)!}{(2n-k)!}$ is a polynomial in $n$ of degree $k$, we can write:
\begin{align}
    \frac{(2n)!}{(2n-k)!} = (2n)(2n-1)\dots(2n-k+1).
\end{align}
Since $z \frac{\partial}{\partial z} z^n = n z^n$, this structure suggests the operator relation:
\begin{align}
    F_k(z) = \left(2z \frac{\partial}{\partial z}\right) \left(2z \frac{\partial}{\partial z}-1\right) \cdots \left(2z \frac{\partial}{\partial z}-k+1\right) F_0(z)
\end{align}
Alternatively, 
\begin{align}
    F_{k+1}(z) = \left(2z \frac{\partial}{\partial z}-k\right) F_k(z). \label{eq:rec.rel}
\end{align}
The general matrix element is then compactly written as:
\begin{empheq}[box=\widefbox]{align}
    \langle \zeta | a^{\dagger k} a^k | \xi \rangle = (1-|\alpha|^2)^{1/4} (1-|\beta|^2)^{1/4} F_k(\alpha\beta^*)
\end{empheq}
\section*{The Squeezing Polynomials and Generating Functions}
The generating function $F_k(z)$ can be constructed from the base case $F_0(z) = (1-z)^{-1/2}$ by applying a differential operator. It is highly useful to factorize $F_k(z)$ into a polynomial part and a singular part. We define the \textbf{squeezing polynomial}, $P_k(z)$, such that:
\begin{equation}
    F_k(z) \equiv \frac{P_k(z)}{(1-z)^{k+1/2}}.
\end{equation}
Substituting this definition into the recurrence relation (\ref{eq:rec.rel}), we can derive a recurrence relation for the polynomials $P_k(z)$ themselves:
\begin{equation}
  \boxed{\quad  P_{k+1}(z) = \Big( (3k+1)z - k \Big) P_k(z) + 2z(1-z) P_k'(z), \quad}
\end{equation}
with the initial condition $P_0(z)=1$. This relation allows for the straightforward generation of any squeezing polynomial. The first few are:
\begin{itemize}
    \item $P_0(z) = 1$
    \item $P_1(z) = z$
    \item $P_2(z) = 2z^2 + z$
    \item $P_3(z) = 6z^3 + 9z^2$
    \item $P_4(z) = 24z^4 + 72z^3 + 9z^2$
    \item $P_5(z) = 120z^5 + 600z^4 + 225z^3$
\end{itemize}
The simple relationship $g^{(k)}(\xi) = P_k(x)/x^k$ (will be derived in \ref{g^k_single}) for a single squeezed state highlights the central role of these polynomials.

\subsection{Explicit Moments and Normalization}

Using this formalism, we can now write the explicit expressions for the moments required in the main text.
\vspace{.5 cm}

\noindent\textbf{The Normalization Constraint $\mathcal{N}_0$:}
The state must be normalized such that $\langle\psi|\psi\rangle=1$. We have:
\begin{equation}
    |\chi|^2 + |\eta|^2 + 2\mathscr{Re}\left[\chi\eta^* \langle\zeta|\xi\rangle\right] = 1.
    \label{eq:norm_main}
\end{equation} 
The overlap term $\langle\zeta|\xi\rangle$ corresponds to the $k=0$ case:
\begin{equation}
    \langle\zeta|\xi\rangle = (1-x)^{1/4}(1-y)^{1/4} F_0(z) = \frac{(1-x)^{1/4}(1-y)^{1/4}}{(1-z)^{1/2}}. \label{xi_zeta_inner}
\end{equation}
This leads to the crucial normalization constraint, which defines the physical parameter space:
\begin{equation}
\boxed{\quad    |\chi|^2 + |\eta|^2 + 2\mathscr{Re}\left[\chi\eta^* \frac{(1-x)^{1/4}(1-y)^{1/4}}{\sqrt{1-z}}\right] = 1. \quad} \label{norm}
\end{equation}

\noindent\textbf{Mean Photon Number $\mathcal{N}_1$:}
Using $F_1(z) = P_1(z) (1-z)^{-3/2}=z(1-z)^{-3/2}$, we find the mean photon number:
\begin{equation}
    \mathcal{N}_1 = |\chi|^2 \frac{x}{1-x} + |\eta|^2 \frac{y}{1-y} + 2\mathscr{Re} \left[ \chi \eta^* (1-x)^{1/4} (1-y)^{1/4} \frac{z}{(1-z)^{3/2}} \right].
\end{equation}

\noindent\textbf{Third-Order Moment $\mathcal{N}_3$:}
Using $F_3(z) = P_3(z) (1-z)^{-7/2}= (6z^3+9z^2)(1-z)^{-7/2}$, we find the third-order moment:
\begin{equation}
    \mathcal{N}_3 = |\chi|^2 \frac{3x^2(2x+3)}{(1-x)^3} + |\eta|^2 \frac{3y^2(2y+3)}{(1-y)^3} + 2\mathscr{Re} \left( \chi \eta^* (1-x)^{1/4} (1-y)^{1/4} \frac{3z^2(2z+3)}{(1-z)^{7/2}} \right).
\end{equation}

\noindent\textbf{$k^\text{th}$-Order Moment $\mathcal{N}_k$:}
\begin{equation}
    \mathcal{N}_k = |\chi|^2 \langle \xi | a^{\dagger k} a^k | \xi \rangle + |\eta|^2 \langle \zeta | a^{\dagger k} a^k | \zeta \rangle + 2\mathscr{Re} \left[ \chi \eta^* \langle \zeta | a^{\dagger k} a^k | \xi \rangle \right].
\end{equation}
Using the generating function formalism, where $\langle \zeta | a^{\dagger k} a^k | \xi \rangle = (1-x)^{1/4}(1-y)^{1/4} F_k(z)$, and the definition of the squeezing polynomial, $F_k(z) = P_k(z)(1-z)^{-(k+1/2)}$, we can rewrite the moment as:
\begin{equation}
  \boxed{\quad   \mathcal{N}_k = |\chi|^2 \frac{P_k(x)}{(1-x)^k} + |\eta|^2 \frac{P_k(y)}{(1-y)^k} + 2\mathscr{Re}\left[\chi \eta^* \frac{(1-x)^{1/4}(1-y)^{1/4} P_k(z)}{(1-z)^{k+1/2}}\right]. \quad} \label{N_k}
\end{equation}
Finally the most general expression for higher order coherence for the Janus state reads
\begin{equation}
  \boxed{\quad   g^{(k)}(0) = \displaystyle \frac{ \displaystyle |\chi|^2 \frac{P_k(x)}{(1-x)^k} + |\eta|^2 \frac{P_k(y)}{(1-y)^k} + 2\mathscr{Re}\left[\chi \eta^* \frac{(1-x)^{1/4}(1-y)^{1/4} P_k(z)}{(1-z)^{k+1/2}}\right]}
  {\Bigg( \displaystyle |\chi|^2 \frac{x}{1-x} + |\eta|^2 \frac{y}{1-y} + 2\mathscr{Re} \left[ \chi \eta^* (1-x)^{1/4} (1-y)^{1/4} \frac{z}{(1-z)^{3/2}} \right] \Bigg)^k}\,
  . \quad} \label{g^k}
\end{equation}

\subsection{Higher-Order Coherence of a Single Squeezed State}

A crucial validation of our general formalism is its ability to recover the known results for its constituent states. We analyze the $k^{\text{th}}$-order coherence, $g^{(k)}$, for a single squeezed state $|\xi\rangle$ by setting $\eta=0$ and $|\chi|=1$ in the general expressions for the Janus state.

The $k^{\text{th}}$ order moment $\mathcal{N}_k = \langle \xi | (a^\dagger)^k a^k | \xi \rangle$ is given by:
\begin{equation}
    \mathcal{N}_k = (1-x)^{1/2} F_k(x),
\end{equation}
where $x = \tanh^2 r$ and $F_k(x) = (1-x)^{-(2k+1)/2} P_k(x)$. The mean photon number is $\mathcal{N}_1 = x/(1-x)$. The coherence function is therefore:
\begin{align}
   g^{(k)}(\xi) = \frac{\mathcal{N}_k}{\mathcal{N}_1^k}= \frac{(1-x)^{1/2} (1-x)^{-(2k+1)/2} P_k(x)}{\left(  \displaystyle \frac{x}{1-x} \right)^k} = \frac{(1-x)^{-k} P_k(x)}{x^k (1-x)^{-k}}.n
\end{align}
This leads to the remarkably simple and powerful relation between the higher-order coherence and the squeezing polynomials:
\begin{align}
    \boxed{\quad g^{(k)}(\xi) =  \displaystyle \frac{P_k(\tanh^2 r)}{\tanh^{2k}r}. \quad} \label{g^k_single}
\end{align}
This formula allows for the direct calculation of any-order coherence function for a single squeezed state, provided the corresponding squeezing polynomial is known.

\vspace{.5 cm}
\paragraph{Example for k=2:}
For second-order coherence, the polynomial is $P_2(x) = 2x^2 + x$. Using the formula above, we recover the well-known result:
\begin{equation}
    g^{(2)}(\xi) = \frac{2x^2+x}{x^2}  = 2 + \coth^2 r = 3 + \frac{1}{\sinh^2 r}.
\end{equation}
\paragraph{Example for k=3:}
For third-order coherence, the polynomial is $P_3(x) = 6x^3 + 9x^2$. This gives:
\begin{equation}
    g^{(3)}(\xi) = \frac{6x^3+9x^2}{x^3}  = 6 + 9\coth^2 r
    = 15 + \frac{9}{\sinh^2 r}.
\end{equation}
This confirms that our general expressions are correct and robust, and it highlights the central role of the squeezing polynomials in determining the photon statistics of these states.

\section*{Asymptotic Analysis of Higher-Order Moments}
We are interested in the regime where $g^{(k)} \to 0$ as a result of interference effects. As evident from Eq.~(\ref{g^k}), for large values of $|\chi| \approx |\eta|$, the function $g^{(k)}$ scales approximately as $|\chi|^{2(1-k)}$, and therefore can approach zero in this limit. However, it is crucial to account for the normalization constraint
\begin{equation}
|\chi|^2 + |\eta|^2 + 2\mathscr{Re}\left[\chi\eta^* \langle\zeta|\xi\rangle\right] = 1.
\end{equation}
The Cauchy–Schwarz inequality implies $\mathscr{Re}\langle\zeta|\xi\rangle \leq |\langle\zeta|\xi\rangle| \leq 1$. If $|\langle\zeta|\xi\rangle| = 1$, then $\ket{\xi} = \ket{\zeta}$, and the Janus state reduces to a single squeezed state with large $g^{(k)}$. Thus, we require $|\langle\zeta|\xi\rangle| < 1$ for genuine interference.

To offset the large values of $|\chi|^2 + |\eta|^2$, the term $2\mathscr{Re}\left[\chi\eta^* \langle\zeta|\xi\rangle\right]$ must be maximally negative, which is achieved when the relative phase $\delta = \pi$. Under this condition, the normalization constraint becomes
\begin{equation}
\left(|\chi| - |\eta|\right)^2 + 2|\chi||\eta|\left[1 - \mathscr{Re}\langle\zeta|\xi\rangle\right] = 1.
\label{eq:constraint_breakdown}
\end{equation}
Since $1 - \mathscr{Re}\langle\zeta|\xi\rangle > 0$, for large amplitudes the constraint can only be satisfied if $1 - \mathscr{Re}\langle\zeta|\xi\rangle$ is very small and $|\chi| - |\eta| \approx \pm 1$.

To achieve $\mathscr{Re}\langle\zeta|\xi\rangle \to 1$, note that the overlap
\begin{equation}
\langle\zeta|\xi\rangle = \frac{(1-x)^{1/4} (1-y)^{1/4}}{(1-z)^{1/2}}
\end{equation}
suggests that, in the limit of small squeezing parameters $r, s \to 0$, we have $x \approx y \approx z \approx 1$. Therefore, the limit of small squeezing must be considered to approach $\mathscr{Re}\langle\zeta|\xi\rangle \to 1$.

\subsection{Small squeezing limit}
To understand the behavior of the Janus state in the limit of small squeezing ($r,s \to 0$), we perform an asymptotic analysis of the $k^{\text{th}}$ order moment, $\mathcal{N}_k$. In the limit of small squeezing, the parameters $x, y, z$ become small. The terms $(1-x)$, $(1-y)$, and $(1-z)$ all approach unity. Therefore, (\ref{N_k}) simplifies to:
\begin{equation}
    \mathcal{N}_k \approx |\chi|^2 P_k(x) + |\eta|^2 P_k(y) + 2\mathscr{Re}\left[\chi \eta^* P_k(z)\right].
\end{equation}
The behavior of this expression is dominated by the lowest-power term in the squeezing polynomial, $P_k(z)$. Let this term be $P_k^{(l_k)} z^{l_k}$, where $l_k = \lceil k/2 \rceil$ is the lowest power of $z$ in the polynomial. Let us also introduce a smallness parameter $\epsilon$ such that $r \sim \epsilon$ and $s \sim \epsilon$. This implies $x \sim \epsilon^2$, $y \sim \epsilon^2$, and $z \sim \epsilon^2 e^{i\Delta}$. Substituting these approximations, the moment $\mathcal{N}_k$ scales as:
\begin{align}
    \mathcal{N}_k &\approx |\chi|^2 P_k^{(l_k)} (r^2)^{l_k} + |\eta|^2 P_k^{(l_k)} (s^2)^{l_k} + 2\mathscr{Re}\left[\chi \eta^* P_k^{(l_k)} (rs e^{i\Delta})^{l_k}\right] \nn\\
    &\approx P_k^{(l_k)} \epsilon^{2l_k} \left( |\chi|^2 + |\eta|^2 + 2\mathscr{Re}\left[\chi \eta^* e^{i l_k \Delta}\right] \right) \nn\\
    &\approx P_k^{(l_k)} \epsilon^{2l_k} \Big( |\chi|^2 + |\eta|^2 + 2|\chi||\eta|\cos(l_k\Delta - \delta) \Big), \label{N_k.approx}
\end{align}
where  $\eta=|\eta| e^{i\delta}$. This final expression is a powerful tool. It shows that the scaling of the $k^{\text{th}}$ order moment in the small-squeezing limit is directly governed by the properties of the squeezing polynomial $P_k(z)$ through its lowest power, $l_k$. Since $l_1=1$ and $P_1^{(1)}=1$, then we have 
\begin{equation}
   \boxed{\quad  g^{(k)}(\psi) \approx \epsilon^{2(l_k-k)} \frac{ |\chi|^2 + |\eta|^2 + 2|\chi||\eta|\cos(l_k\Delta - \delta) }{\Big[ |\chi|^2 + |\eta|^2 + 2|\chi||\eta|\cos(\Delta - \delta) \Big]^k}. \quad} \label{small_r,s}
\end{equation}
It is clear from Eq.~(\ref{small_r,s}) that given $|\chi|$ and $|\eta|$ are not large, then,  $g^{(k)} \propto \epsilon^{2(l_k-k)} \to \infty$. So we conclude to find out the limit where $g^{(k)}$ reaches zero, we have to consider large $|\chi|$ and $|\eta|$. 

For this to hold as $|\chi|, |\eta| \to \infty$, we must choose the phase to be maximally destructive, $\delta=\pi$. This leads to the condition $(|\chi|-|\eta|)^2 \approx 1$, or $|\chi| \approx |\eta| \pm 1$. Using the relation 
\begin{align}
    |\chi|^2+|\eta|^2+2|\chi||\eta|\cos\theta = (|\chi|-|\eta|)^2+2|\chi||\eta|(1+\cos\theta) \approx 1+2|\chi||\eta|(1+\cos\theta),
\end{align}
and using  $\delta=\pi$ in (\ref{small_r,s}), the coherence function becomes:
\begin{equation}
   \boxed{\quad  g^{(k)}(\psi) \approx \epsilon^{2(l_k-k)} \frac{ 1+2|\chi||\eta|(1-\cos(l_k\Delta)) }{\Big[ 1+2|\chi||\eta|\big(1-\cos(\Delta)\big) \Big]^k}. \quad} \label{small_r,s_large_eta}
\end{equation}
The behavior of the system is critically dependent on the interference phase $\Delta$. We consider two separate cases in the following.

\subsection{Case 1: Trivial Interference ($\Delta=0$)}

In the trivial interference case, the coherence function diverges in the small squeezing limit, as $g^{(k)} \propto \epsilon^{2(l_k-k)} \to \infty$ regardless of the amplitudes.

\subsection{Case 2: Non-Trivial Interference ($\Delta \neq 2n\pi$)}
The non-trivial interference case ($\Delta \neq 0$) by considering two sub-cases depending on the value of $l_k\Delta$.

\paragraph{Sub-case I: $1-\cos(l_k\Delta) = 0$.}
This special condition occurs if $l_k\Delta$ is a multiple of $2\pi$. In this scenario, the general scaling becomes:
\begin{equation}
    g^{(k)}(\psi) \propto \frac{1}{(\epsilon |\eta|^2)^k} \approx r^k \to 0, \quad \text{for } |\eta| \approx \frac1{\epsilon}\approx \frac1{r}.
\end{equation}
where for even $k$, $l_k=k/2$. This shows that even when the primary interference term is zero, the state still exhibits strong antibunching. Moreover, for the odd $k$, $l_k=(k+1)/2$, and hence
\begin{equation}
    g^{(k)}(\psi) \propto \frac{1}{\epsilon^{k-1} |\eta|^{2k}} \approx r^{k+1} \to 0, \quad \text{for } |\eta| \approx \frac1{\epsilon}\approx \frac1{r}.
\end{equation}

\paragraph{Sub-case II: $1-\cos(l_k\Delta) \neq 0$.}
When $\Delta \neq 0$, the large-$|\eta|$ limit is permitted by the normalization constraint. For large $|\eta|$, the `1` in the brackets of the $g^{(k)}$ expression is negligible, and the expression simplifies to a scaling law:
\begin{equation}
    g^{(k)}(\psi) \propto \frac{\epsilon^{2l_k} |\eta|^2 (1-\cos(l_k\Delta))}{\epsilon^{2k} |\eta|^{2k} (1-\cos\Delta)^k} = \epsilon^{2(l_k-k)} |\eta|^{2(1-k)} \frac{1-\cos(l_k\Delta)}{(1-\cos\Delta)^k}.
\end{equation}
For odd $k$, we have $l_k=\frac{k+1}{2}$, which gives the specific scaling:
\begin{equation}
   \boxed{\quad g^{(k)}(\psi) \approx  \frac{1}{(\epsilon |\eta|^2)^{k-1}}  \approx  r^{k-1},  \quad \text{for } |\eta| \approx \frac1{\epsilon}\approx \frac1{r}. \quad}
\end{equation}
For instance, for $k=3$, this demonstrates that $g^{(3)} \to 0$ under the condition that $\epsilon |\eta|^2 \gg 1$. For a general even $k$, we have $l_k = k/2$. The scaling is:
\begin{equation}
    \boxed{\quad g^{(k)}(\psi) \propto  \epsilon^{-k} |\eta|^{2(1-k)}\approx  r^{k-2},  \quad \text{for } |\eta| \approx \frac1{\epsilon}\approx \frac1{r}. \quad}
\end{equation}
This shows that $g^{(k)} \to 0$ under the slightly more restrictive condition, i.e.,  $\epsilon |\eta|^{2(k-1)/k} \gg 1$.

\section{Some numerical numbers}

The theoretical analysis above reveals the rich parameter dependence and tunability of higher-order coherence in the Janus state. To illustrate these predictions concretely, Table~\ref{tab:g3_examples} provides representative numerical examples covering a broad range of physical regimes. The table highlights how different choices of squeezing, phases, and amplitudes allow one to access either strong photon antibunching or pronounced bunching. These results underscore the extreme sensitivity of $g^{(3)}$ to quantum interference, offering a direct bridge between the analytical scaling laws and experimentally relevant parameter sets.

\begin{table*}[t]
\caption{Numerical examples demonstrating the extreme tunability of the third-order coherence, $g^{(3)}$, for the Janus state. The table shows the required superposition amplitudes ($|\chi|, |\eta|$) to achieve a target $g^{(3)}$ for various squeezing parameters ($r, s$) and phase configurations ($\Delta, \delta$). The results are grouped to highlight the different physical regimes, from strong suppression (antibunching) to extreme bunching.}
\label{tab:g3_examples}
\centering
\begin{ruledtabular}
\begin{tabular}{lccccccc}
\textbf{Scenario Description} & \textbf{$\boldsymbol{\Delta}$} & \textbf{$\boldsymbol{\delta}$} & \textbf{$\boldsymbol{r}$} & \textbf{$\boldsymbol{s}$} & \textbf{$\boldsymbol{|\eta|}$} & \textbf{$\boldsymbol{|\chi|}$} & \textbf{Resulting $\boldsymbol{g^{(3)}}$} \\
\hline
\multicolumn{8}{c}{\textit{\textbf{Group 1: Strong Suppression Regime (Three-Photon Antibunching)}}} \\
\hline
Symmetric Suppression & $\pi$ & $\pi$ & 0.100 & 0.100 & 3.7268 & 4.5425 & 0.01 \\
Symmetric Suppression & $\pi$ & $\pi$ & 0.250 & 0.250 & 1.9536 & 2.5934 & 0.10 \\
Asymmetric Suppression & $\pi$ & $\pi$ & 0.100 & 0.150 & 1.1369 & 2.0996 & 0.50 \\
\hline
\multicolumn{8}{c}{\textit{\textbf{Group 2: Mid-Range Bunching Regime}}} \\
\hline
Asymmetric Anti-Symmetric & $\pi$ & $\pi$ & 0.100 & 0.200 & 0.5642 & 1.5449 & 5.00 \\
Asymmetric Anti-Symmetric & $\pi$ & $\pi$ & 0.100 & 0.250 & 0.6121 & 1.5831 & 10.00 \\
\hline
\multicolumn{8}{c}{\textit{\textbf{Group 3: Extreme Bunching Regime (Single Squeezed State Limit)}}} \\
\hline
Single Squeezed State & N/A & N/A & 0.100 & 0.100 & 0 & 1 & 912.01 \\
Single Squeezed State & N/A & N/A & 0.050 & 0.050 & 0 & 1 & 3612.00 \\
Single Squeezed State & N/A & N/A & 0.010 & 0.010 & 0 & 1 & 90012.00 \\
\end{tabular}
\end{ruledtabular}
\end{table*}

\end{document}